\documentclass[aps,pra,preprint,showkeys]{revtex4-2}
\usepackage{amsmath, latexsym, amssymb, amsbsy}
\usepackage{graphicx}
\usepackage{color}
\usepackage{latexsym,amsmath,amssymb,amsbsy,graphicx}
\usepackage{epstopdf}
\DeclareGraphicsExtensions{.eps}
\usepackage{bm}

\DeclareMathOperator{\rot}{rot}

\def\bea{\begin{eqnarray}}
\def\eea{\end{eqnarray}}
\def\be{\begin{equation}}
\def\ee{\end{equation}}
\def\nn{\nonumber}
\def\d{{\rm d}}

\def\be{\begin{eqnarray}}
\def\ee{\end{eqnarray}}

\begin{document}
\title{Angular momentum transfered by the field of a moving point charge}

\author{Vladimir Epp}
 \altaffiliation{Department of Theoretical Physics, Tomsk State Pedagogical University, 634061 Tomsk, Kievskaya 60, Russia.}
 \email{epp@tspu.edu.ru}
\author{Ulyana Guselnikova}%
\affiliation{Department of Theoretical Physics, Tomsk State Pedagogical University, Russia, 634061 Tomsk, Kievskaya 60, Russia.}
\author{Irina Kamenskaya}%
\affiliation{General Physics Department, Tomsk State Pedagogical University, Russia, 634061 Tomsk, Kievskaya 60, Russia.}
\date{\today}
\begin{abstract}
The flux of angular momentum of electromagnetic field of an arbitrarily moving point charge is investigated. General equations are obtained for the transfer of angular momentum at arbitrary distance from the charge, and corresponding equations in the far-field approximation. An explicit expression is obtained for the flux of angular momentum in the wave zone in terms of coordinates, velocity, and acceleration of the charge. The torque is calculated, that would act on an object if it absorbed all the radiation incident on it. It is shown that this torque is proportional to the curl of the stress tensor of the electromagnetic field; in the far field approximation the torque is proportional to the curl of the Poynting vector.
\end{abstract}

\keywords{orbital angular momentum; torque; electromagnetic field; charged particle}

\maketitle
\section{Introduction}
It is well known that  electromagnetic field of a point charge carries angular momentum \cite{Jackson_Cl_El}. The angular momentum is usually devided into a spin part associated with polarization  and an orbital part associated with helical wave front.
A light beam with a  helical wave front is usually referred to as a twisted light.  The first theoretical and experimental research of  twisted light or vortex radiation, were devoted to laser radiation modified by astigmatic optical system, numerically computed holograms \cite{Allen1992, Padgett2004}, or microscopic spiral phase plates. 
These publications have stimulated  extensive studies of vortex optical beam.  Further  extensive bibliography on the history in this field  can be found  in the recent review \cite{Barnett2017}.
Interest in  vortex radiation quickly spread over different areas of physics: transfer of information,  interaction with atoms,  high-energy particle collision and radiation processes. 
The vortex light beams  have opened a wide range of applications, such as spatial optical trapping of atoms or microscopic objects,   phase-contrast microscopy, and nano- or micro-scale physics \cite{Barnett2017,Andrews2012book}. Various methods are being developed to register the orbital angular momentum of radiation \cite{Chen2015, Li2019}.

High energy photons carrying angular momentum can be   emitted by the vortex  beams of charged particles.
 In recent years  attention to the X-ray vortices produced by high energy particles has been boosted by the interest to  microscopy and spectroscopy in the atomic and nanometric scale. 
 Radiation in the X-ray range carrying the angular momentum was obtained by converting an x-ray beam \cite{Kohmura2009} and by use of a helical undulator \cite{Bahrdt2013}. Various schemes of twisted photon beam production in undulators \cite{ Bordovitsyn2012,Matsuba_2018, EppGuselnikova2019} and free electron lasers  \cite{Hemsing2011} have been proposed. 
 Cherenkov radiation and transition radiation emitted by vortex electrons was studied theoretically \cite{KarlovetsPhysRev, KarlovetsPhysRevLett, Konkov2014}.
 Radiation of high energy charged particles  channeled in the solid and liquid crystals has been studied theoretically in recent papers \cite{Abdrashitov2018, EppJanz2018,Bogdanov_2021, Epp2021}.
The  X-ray vortex radiation has found numerous applications both in classical and quantum optics condensed matter, high energy physics, optics, etc. (see  the review \cite{Hernandez-Garcia2017} and references therein).

Despite active experimental research, there is a lack of theoretical studies on transfer of angular momentum by the field of an arbitrarily moving charge. The available theoretical papers in this area cover the angular momentum of radiation only in some special cases, such as synchrotron radiation, radiation in crystals, or polarization radiation.

This paper aims to fill this gap. In section \ref{II} we obtain a general expression for the flux of angular momentum through a unit area in terms of the stress tensor of the field of a point charge. 
In section \ref{III}, we study the flux of angular momentum in the far-field approximation.  Since the angular momentum essentially depends on the specific coordinate system, we consider two cases -- when the coordinate origin is relative far from the charge, and when the distance between the charge and the coordinate origin is much less than the distance to the observation point. Approximate expressions for the flux of angular momentum of a nonrelativistic charge are obtained in section \ref{IV}. Section \ref{V} is devoted to study of the torque exerted on an area element due to the electromagnetic field. The obtained expressions were applied to calculate the angular momentum flux and the torque in the field of a rotating dipole  in section \ref{VI}. Finally, section \ref{VII} is devoted to  discussion of the results obtained.

\section{\label{II}The flux of angular momentum in the field of a point charge}
 
 The angular momentum tensor $L^{\mu\nu}$ of the electromagnetic field  is defined by the energy-momentum tensor  $T^{\mu\sigma}$ as \cite{Landau_II}
\be\label{L144'}
L^{\mu\nu}=\int (x^\mu{T}^{\nu\sigma}-x^\nu T^{\mu\sigma})\d S_\sigma, 
\ee
where $\d S_\sigma$ is the vector equal in magnitude to the area of a  hypersurface element and normal to this element. The spatial components of the energy-momentum  tensor form a three-dimensional stress tensor
\[
\sigma_{ik}=\frac{1}{4\pi}\left[-E_i E_k-H_i H_k+\frac{1}{2}\delta_{ik}(E^2+H^2)\right],
\]
with $E_i$ and $H_i$ being the components of  electric and magnetic fields, respectively, $\delta_{ik}$ is the Kronecker symbol. The time components determine the energy and momentum density of the electromagnetic field 
\[
T^{00}=\frac{E^2+H^2}{8\pi},\quad T^{0i}=\frac 1c P_i,\quad \bm P=\frac{c}{4\pi}(\bm E\times \bm H),
\]
$\bm P$ is the Poynting vector.

Passing to three-dimensional notation, we introduce a three-dimensional angular momentum vector with components $L_i=\frac{1}{2}e_{ijk}L^{jk}$, $e_{ijk}$ is the unit antisymmetric symbol. The flux of the $i$-th component of the vector $\bm L$ through the unit area orthogonal to the $k$-th axis is determined by the three-dimensional tensor \cite{Landau_II,Sokolov1991}
\be\label{gik}
g_{ik}=e_{ijm}x_j\sigma_{mk}=\frac{1}{4\pi}\left[- (\bm r  \times\bm E )_i E_k-(\bm r\times \bm H)_i H_k+\frac 12(E^2+H^2)e_{ijk}r_j \right].
\ee
The flux of the angular momentum of the field through an arbitrarily oriented area $ \bm {\d s} $ is
\be\label{Lds11}
\frac{\d\bm L}{\d t}=\frac{1}{4\pi}\left[- (\bm r  \times\bm E )(\bm E\,\bm{\d  s})-(\bm r\times \bm H)(\bm H\,\bm{\d  s})+
\frac 12(E^2+H^2)(\bm r\times\bm{\d  s})\right].
\ee
If the area $ \bm {\d s}$ is orthogonal to the vector $ \bm r $, then the flux  is 
\be\label{Lds}
\frac{\d\bm L}{\d t}=\frac{r}{4\pi}\left[(\bm E \times \bm n )(\bm E\,\bm  n)+(\bm H\times \bm n)(\bm H\,\bm n)\right]\d s.
\ee
Here $\bm n=\bm r/r$. 
 In the spherical coordinate system  $r,\,\theta,\,\varphi$ with unit vectors $\bm e_r,\bm e_\theta,\bm e_\varphi$  the angular momentum flux in radial direction is given by
\be\label{Lds-sp}
\frac{\d\bm L}{\d t\d s}=r\left(\sigma_{12}\bm e_\varphi- \sigma_{13} \bm e_\theta\right).
\ee

The electric and magnetic fields of an arbitrary moving charge at a point $\bm r$ and at a time moment $t$   are  \cite{Landau_II}
\begin{align}\label{EH}
\bm E&=\bm E_1+\bm E_2,\quad \bm H=(\bm R\times \bm E)/R,\\
\label{ee}
\bm E_1&=\frac{eR^2\bm \kappa}{c(R-\bm \beta\bm R)^3},\quad
\bm E_2=\frac{e(1-\beta^2)(\bm R-R\bm\beta)}{(R-\bm \beta\bm R)^3},\\
\bm \kappa&=[\bm R\times[(\bm R-R\bm\beta)\times\bm{\dot{\beta}}]]/R^2,\nn
\end{align}
where   $\bm R=\bm r-\bm r'$ is the vector from the charge position to the point  $\bm r$, $\bm r'=\bm r'(t')$ is the charge position  at the retarded  time moment $t'=t-R/c$, $c$ is the speed of light, $e$ is the charge, $\bm \beta=\bm v/c,\, \bm v=\bm v(t')$ is the particle velocity, and the dot denotes the time derivative.
\section{Flux of angular momentum in the far field approximation\label{III}}
Eqs. (\ref{gik}) -- (\ref {ee}) determine the flux of angular momentum of the electromagnetic field at any distance from the charge. To find the angular momentum carried by the radiation of the charge, we obtain approximate equations for the angular momentum at large distance. The field $ \bm E_1 $ decreases with distance as $ 1 /R $, and the field $ \bm E_2 $ as $ 1 /R ^ 2 $. At large distances from the charge, where $ R \gg c \dot \beta $, the electric field $ \bm E_1 $ and the corresponding magnetic field prevail. Therefore, when calculating the intensity of radiation we can neglect the field $ \bm E_2 $.

\begin{figure}[ht]\center
\includegraphics[width=7cm]{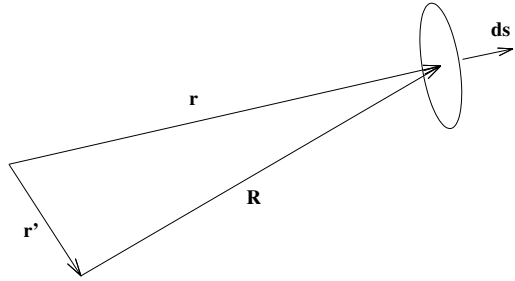}
\caption{Notations used.}
\label{(1)}
\end{figure}

\begin{figure}[ht]\center
\includegraphics[width=6cm]{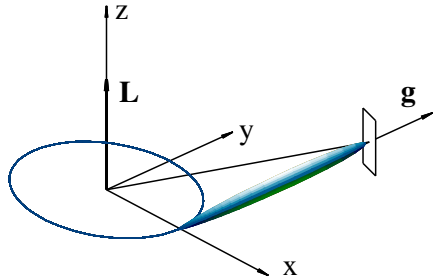}
\caption{Synchrotron radiation. The main part of the angular momentum transferred to the target is due to the radiation pressure  $ \bm g $ on the target which causes   the  torque relative to the $ z $ axis.}
\label{(2)}
\end{figure}

Calculation of the angular momentum of radiation has one important difference from the calculation of the  intensity of radiation  -- the angular momentum depends significantly on the choice of the coordinate system. When calculating the  intensity of radiation, two conditions are usually assumed to be satisfied: i) $ R \gg c \dot \beta $, then the field $ \bm E_2 $ can be neglected, and ii) $ R, r \gg r '$ (see Fig. \ref {(1)}) and then you can put $ \bm r \approx \bm R $.
However, in practice, there are cases when the second condition is not met. For example, when calculating the intensity of synchrotron radiation, it is convenient to choose the origin at the center of a circular orbit, but measurements are carried out at a distance comparable to or even less than the orbital radius, as, for example, in Fig. \ref {(2)}. Nevertheless, the approximation i) $ R \gg c \dot \beta $ (far field or wave zone approximation) is sufficient to calculate the radiation intensity. 

A completely different situation arises when calculating the angular momentum of radiation. Two substantially different situations should be distinguished here:
i) the target is in the wave zone ($ R \gg c \dot \beta $), but the distances $ R $ and $ r '$ are comparable and ii) the target is in the wave zone and
$ R, r \gg r '$. Let's consider these cases separately.
\subsection{Angular momentum in the wave zone when $r'\sim r,R$}\label{3_1}
In this case, we can neglect the field $ \bm E_2 $ in  Eq. (\ref {Lds}), and put $ \bm E \approx \bm E_1 $. Then the vectors of the electric and magnetic fields are mutually orthogonal and orthogonal to the direction of radiation $ \bm N = \bm R / R $. If we put in Eq. (\ref {Lds}) $ \bm H = \bm N \times \bm E, \, (\bm E \bm N) = 0 $, then it can be transformed to the form
\be\label{Lds-n}
\frac{\d\bm L}{\d t\d s}=\frac{r}{4\pi} E^2(\bm n \times \bm N )(\bm n \bm N)\d s=\frac{1}{c} (\bm r\times \bm P )(\bm N \bm{n}),
\ee
Thus, if the distances $ r ', \, r, \, R $ are comparable in magnitude, then the flux of angular momentum in the radiation field can be considered as a mechanical torque due to the radiation pressure. The multiplier $ (\bm n \bm N) $ in Eq. (\ref {Lds-n}) takes into account that the area
$ \bm {\ d s} = \bm n \d s$ is oriented at an angle to the direction of radiation $ \bm N $. In this case, the angular momentum of radiation can be definitely considered as the orbital angular momentum.
\subsection{Angular momentum in the wave zone when $ R, r \gg r '$}
Let us consider the case when the charge moves in a region whose dimensions are much less than the distance $ R $ and we are interested in the angular momentum of the field relative to some point lying in the region of motion of the charge ($ r '\ll R $). Then the scalar product $ (\bm E \, \bm n) $ in the Eq. (\ref {Lds}) becomes small and decreases with distance as $ 1 / R ^ 2 $. In this case, the contribution of the $ \bm E_2 $ field should also be taken into account. In the limit $ r \to \infty $,  Eq. (\ref {Lds}) takes the form
\be\label{Lds1}
\frac{\d\bm L}{\d\Omega\d t}=\frac{r^3}{4\pi}\left[(\bm E_1 \times \bm n )(\bm E_2\,\bm  n)+(\bm H_1\times \bm n)(\bm H_2\bm n)\right],
\ee
where $ \bm H_i = (\bm R \times \bm E_i) / R $. Here each term decreases as $ 1 / r ^ 3 $ as $ r \to \infty $.

This expression can also be expressed in terms of the Poynting vector. If we make the substitution $ \bm H_1 = \bm n \times \bm E_1 $ in  Eq. (\ref {Lds1}), then we get
\be\label{Lds2}
\frac{\d\bm L}{\d s\,\d t}=\frac{r}{4\pi}\left[\bm E_1(\bm H_2\bm n )-\bm H_1(\bm E_2\bm n)-\bm n (\bm H_2\bm n)(\bm E_1\bm n)\right].
\ee
The first two terms decrease with distance as $ 1 / r ^ 3 $, and the last term as $ 1 / r ^ 4 $. Therefore, it can be neglected.
On the other hand, in the considered approximation
\be\label{pper}
\bm n\times \bm P=\frac{c}{4\pi}[\bm E_1(\bm n\bm H_2)-\bm H_1(\bm n\bm E_2)].
\ee
Hence, the flux of angular momentum in the wave zone  in approximation  $ R, r \gg r '$ can be calculated through the Poynting vector
\be\label{Lds-n1}
\frac{\d\bm L}{\d s\,\d t}=\frac{1}{c} (\bm r\times \bm P).
\ee
In the approximation $ r '\ll r $, this equation. in essence. coincides with  Eq. (\ref {Lds-n}), since in the expansion of the scalar product $ (\bm N \bm {\d s}) $ in powers of $ r '/ r $ the first term is unit. As expected, the flux of angular momentum in the wave zone of a point charge is described by the well-known equation for the flux of angular momentum of electromagnetic radiation \cite {Jackson_Cl_El, Panofsky1962}.

The flux of the modulus of the angular momentum of  radiation can be related to the radiation intensity. Let us denote by $ \alpha $ the angle between the vectors $ \bm n $ and $ \bm P $, and by $ I $ -- the radiation intensity (energy per unit time). Then
\be\label{Ldssin}
\frac{\d|\bm L|}{\d s\,\d t}=\frac{1}{c} r\sin\alpha\frac{\d I}{\d s}.
\ee

Let us find an explicit expression for the angular momentum flux in terms of the charge position, its velocity and acceleration.
Substituting $ \bm R = \bm r- \bm r '$ into Eqs. (\ref {EH}), (\ref {ee}) and (\ref {Lds1}), and keeping the largest term in the expansion in powers of $ 1 / r $, we obtain
\be\label{vec}
\frac{\d\bm L}{\d\Omega\d t}=\frac{e^2}{4\pi c (1-\bm\beta\bm n)^5}\left[ (1-\beta^2)(\bm\kappa\times \bm n)+\frac{\kappa^2(\bm r'\times\bm n)}{c(1-\bm\beta\bm n)}\right].
\ee

Now, the vector $ \bm \kappa $  contains only the main term of the expansion in powers of $ 1 / r $:
\be\label{kap}
\bm \kappa=[\bm n\times[(\bm n-\bm\beta)\times\bm{\dot{\beta}}]].
\ee

This equation determines the angular distribution of the angular momentum carried by  radiation of an arbitrarily moving charge. The first term in  Eq. (\ref {vec}) does not depend on the choice of the origin of coordinate system, while the second one linearly depends on the position vector of the charge. In this sense, the first term can be associated with the spin of radiation, and the second one with the orbital angular momentum.
 The classical theory, in principle, does not distinguish the spin and orbital parts of the total angular momentum. Nevertheless, there is an active discussion on the issue of dividing the angular momentum of  electromagnetic field into orbital and spin parts. See, for example, \cite{Barnett2016, Bordovitsyn2012,Teitelboim1980} and references therein. We would not plunge into this discussion here. Therefore, we denote the studied total angular momentum just as the angular momentum of radiation.

\section{Non-relativistic approximation\label{IV}}
In the non-relativistic approximation $ \beta \ll 1 $. Assuming in Eq. (\ref {vec}) $ \beta = 0 $ and taking into account that for $ \beta = 0 $ the vector product $ (\bm \kappa \times \bm n) = (\bm n \times \bm {\dot {\beta}}) $, in the first approximation in $ \beta $ we have
\be \label {ner1}
\frac {\d \bm L} {\d \Omega \d t} = \frac {e ^ 2} {4 \pi c} (\bm n \times \bm {\dot {\beta}}).
\ee
Hence, the angular momentum of the radiation  is orthogonal to the vectors $ \bm n $ and $ \bm {\dot {\beta}} $.
In the spherical coordinate system $ (r, \theta, \varphi) $ with the $ z $ axis directed along $ \bm {\dot {\beta}} $ and the polar angle $ \theta $, there remains only one component of the vector $ \bm L $ corresponding to the azimuth angle $ \varphi $
\be\label{ner2}
\frac{\d L_\varphi}{\d\Omega\d t}=\frac{e^2\dot\beta}{4\pi c } \sin\theta.
\ee
Obviously, if a particle moves within a limited region of space, then the  angular momentum transferred by the particle's field is time dependent. The linear dependence on $ \dot \beta (t) $ vanishes the average  angular momentum to zero. Actually, Eq. (\ref {ner2}) describes the exchange of angular momentum between the near and far zones and, therefore, does not represent emission of angular momentum. Moreover, the angular momentum of the radiation integrated over the solid angle is in this approximation zero.

In order to calculate the nonzero mean value of the angular momentum flux, it is necessary to take into account the next term in the expansion in $ \beta $. As a result of the expansion up to $ \beta ^ 2 $, we obtain
\be\label{vecl12}
\frac{\d\bm L}{\d\Omega\d t}=\frac{e^2}{4\pi c}\left[ (\bm n\times \bm{\dot{\beta}})(1+4\bm\beta\bm n)+(\bm n\times \bm \beta)(\bm n\bm {\dot\beta})\right].
\ee

Let us find the time-average angular momentum flux. When averaging over time $ t $, one should keep in mind that the right-hand side of the last expression depends on $ t '$, and $ \d t = (1- \bm \beta \bm n) \d t' $. So
\be\label{vecln}
\left\langle\frac{\d\bm L}{\d\Omega\d t}\right\rangle=\frac{e^2}{4\pi c}\frac 1T\int\limits_0^T\left[ (\bm n\times \bm{\dot{\beta}})(1+4\bm\beta\bm n)+(\bm n\times \bm \beta)(\bm n\bm {\dot\beta})\right](1-\bm\beta\bm n)\d t'.
\ee
The average of the term linear in $ \bm {\dot \beta} $ is zero, and averaging the remaining term gives
\be\label{vecl13}
\left\langle\frac{\d\bm L}{\d\Omega\d t}\right\rangle=\frac{e^2}{4\pi c}\left\langle 3 (\bm n\times \bm{\dot{\beta}})(\bm\beta\bm n)+(\bm n\times \bm \beta)(\bm n\bm {\dot\beta})\right\rangle.
\ee
Integrating this expression over the solid angle, we obtain the total angular momentum emitted per unit time
\be\label{vecl14}
\left\langle\frac{\d\bm L}{\d t}\right\rangle=\frac{2e^2}{3 c}\left\langle  \bm\beta\times \bm{\dot{\beta}}\right\rangle.
\ee
The vector $ (\bm \beta \times \bm {\dot {\beta}}) $ indicates the direction of the instantaneous mechanical angular momentum of the particle.
The last expression up to sign coincides with the well-known equation for the loss of angular momentum of a charged particle due to radiation friction \cite {Landau_II}.
\section{Torque exerted by the field\label{V}}

By definition, the angular momentum of  electromagnetic field depends on the choice of the coordinate system. Under certain conditions, as discussed in section \ref {3_1}, radiation pressure can make a significant contribution to angular momentum.
For example, the orbital angular momentum of a photon of synchrotron radiation in an accelerator (Fig. \Ref {(2)}) is equal to the product of the photon momentum by the orbital radius $a$
\[
L=\frac{\hbar\omega}{c}a\sim \gamma^3\hbar,
\]
where $ \omega \sim \gamma ^ 3 c / a $ is the photon frequency, $ \gamma = (1- \beta ^ 2) ^ {- 1/2} $ is the relativistic factor. In modern accelerators, this value can reach gigantic values.
In this case, as in many others, the angular momentum of the radiation  relative to the geometric center of the trajectory is of interest. But if we investigate the radiation of an unknown source, then the position of this geometric center is not determined. Or the source of radiation is so far away that the entire area of radiation shrinks to a point, as is usually the case in astronomy. In these cases, Eq. (\ref {vec}) has little sense. Only the  integral over the solid angle  is important, because it determines the rate of loss of angular momentum by the charge due to radiation friction.

However, the `vortex radiation'  has a property that can theoretically be measured without being bound to any particular coordinate system. 
To date, a large number of experiments have been carried out in which a vortex laser beam drives the microparticles  into rotation \cite {Zhao:09, ONeil2002, Volke2002, Garces_Chavez2003}.
The first experiments on the capture of particles in traps and their rotation were carried out at the end of the last century \cite {He1995, Simpson1997} and are still being carried out \cite {ONeil2002, Volke2002, Garces_Chavez2003}. A detailed description of these works is given in recent review \cite {Bruce2021}.

Obviously, the component of  angular momentum parallel to the normal to an area is not transferred through this area, 
since the diagonal elements of the tensor $ g_ {ij} $ are  zero. However, if the field of the stress tensor is not uniform and the area is of finite diminsion, this is quite possible.

\begin{figure}[ht]\center
\includegraphics[width=7cm]{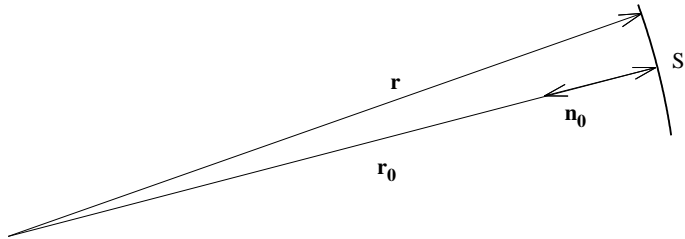}
\caption{The flux of the normal component of the angular momentum of the field through the area $ S $.} 
\label{disk1}
\end{figure}

Consider a small area $ S $ oriented orthogonel to the radius vector passing through the center of the area. The smallness of the area means that its dimensions are much less than the distance to the origin of coordinate system relative to which the angular momentum is calculated.
 Let's denote the coordinate of the center of the area through $ \bm r_0 $ as shown in Fig. \ref {disk1}.
 The unit normal  to the area is $ \bm n_0 = - \bm r_0 / r_0 $. 
  Let's find the torque acting on the area due to the total flux of the angular momentum through the area. The physical meaning of the components of the tensor $ \sigma_ {ij} $ is that they represent forces acting per unit area. The diagonal elements represent pressure, and the off-diagonal element $\sigma_{ij}$ is shear acting in the $ i $-th direction on a unit area with the $ j $-th normal.
 
\begin{figure}[ht]\center
\includegraphics[width=3.8 cm]{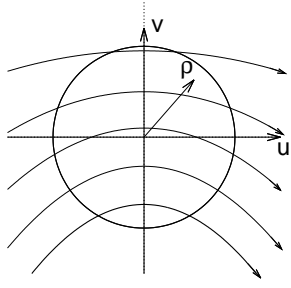}
\caption{Lines of force of the vector $ g_ {i1} $ on the area $ S $.}
\label{circ}
\end{figure}

 Fig. \ref {circ} schematically shows the lines of the shear force  vector $ \sigma_ {i1} $ on the area orthogonal to the radius vector $ \bm r_0 $. If the field of shears is not uniform,  they produce a torque acting on the area. 
  The flux of the normal component of the angular momentum  through the area is given by integral
\be\label{lnt}
 \frac{\d L_n}{\d t}=\int n_{0i}g_{i1}(\bm r)\d s=\int n_{0i}e_{ijm}x_j\sigma_{m1}(\bm r)\d s.
 \ee
Here $ g_ {ij} (\bm r) $ is the tensor (\ref {gik}), index 1 corresponds to the radial coordinate. Obviously, the flux of the normal component of the angular momentum through the center of the area is zero, since $ n_ {0i} e_ {ijm} x_ {0j} \sim \bm n_ {0} \times \bm r_0 = 0 $. However, as can be seen from Fig. \ref {disk1}, at a distance from the center, the cross product $ \bm n_ {0} \times \bm r $ is no longer  zero.
 
 Let us expand the components of the tensor $ \sigma_ {m1} (\bm r) $ in a Taylor series in the vicinity of the center. Further on, we use the spherical coordinate system $ (x_1, x_2, x_3) = (r, \theta, \varphi) $. Let us introduce on the area mutually orthogonal coordinates $ u, \, v $ such that the coordinate lines of $ u $ and $ v $ coincide with the coordinate lines of $ \theta $ and $ \varphi $, respectively.
  Then
\[
\sigma_{m1}(\bm r)=\sigma_{m1}(\bm r_0)+\left( u\frac{\partial \sigma_{m1}(\bm r)}{\partial u}\bigg\rvert_0+v\frac{\partial \sigma_{m1}(\bm r)}{\partial v}\bigg\rvert_0\right)
\] 
The vertical bar means that the  value of derivatives is taken at the center. Next we substitute this in Eq. (\ref {lnt}). The vector $ \bm n_0 $ has coordinates $ n_ {0i} = (n_ {01}, - u / r, -v / r). $ For the sake of simplicity, assume that the area is a disk. Using polar coordinates on the surface of the disk
\[
u=\rho\cos\psi,\quad v=\rho\sin\psi
\]
we transform the integral (\ref {lnt})  to the form
\be
 \frac{\d L_r}{\d t}=\pi \left(\frac{\partial \sigma_{31}(\bm r)}{\partial u}\bigg\rvert_0-\frac{\partial \sigma_{21}(\bm r)}{\partial v}\bigg\rvert_0\right)
 \int \rho^3\d \rho=\frac{1}{4\pi}S^2\left(\frac{\partial \sigma_{31}(\bm r)}{\partial x_2}\bigg\rvert_0-\frac{\partial \sigma_{21}(\bm r)}{\partial x_3}\bigg\rvert_0\right),
\ee
where $ S $ is the area of the disk. Thus, the flux of the normal component of the angular momentum through the area is proportional to the square of its area and to the curl  of $ \sigma_ {ij} $, taken with respect to the first index.

If the expression in parentheses is denoted by $ \Omega_n $
\be\label{omn}
\Omega_n=\left(\frac{\partial \sigma_{31}}{\partial x_2}-\frac{\partial \sigma_{21}}{\partial x_3}\right)\bigg\rvert_0,
\ee
then the torque acting on the area orthogonal to the radius vector is written as
\be \label {rott}
 \frac {\d L_n} {\d t} = \frac {1} {4 \pi} S ^ 2 \Omega_n.
\ee
The non-diagonal elements of the tensor $ \sigma_ {ij} $ decrease with distance as $ 1 / r ^ 3 $, the curl of these components decreases as $ 1 / r ^ 4 $, therefore the torque acting on the area of finite dimensions decreases with distance as $ 1 / r^ 4 $, as for example in Eq. (\ref{r-4}). This is associated with difficulties in measuring the torque of radiation at large distances.

Obviously, the calculations performed can be generalized to areas orthogonal to other coordinate lines.
 The curl of the stress tensor $ \sigma_ {ij} $ is a tensor of the second rank
\be
R_{ij}=e_{kli}\frac{\partial \sigma_{lj}}{\partial x_k}.
\ee
From the diagonal elements of this tensor, we can form the vector
\be\label{bm-Om}
\bm\Omega=\left(\frac{\partial \sigma_{31}}{\partial x_2}-\frac{\partial \sigma_{21}}{\partial x_3},
\frac{\partial \sigma_{12}}{\partial x_3}-\frac{\partial \sigma_{32}}{\partial x_1},
\frac{\partial \sigma_{23}}{\partial x_1}-\frac{\partial \sigma_{13}}{\partial x_2}\right),
\ee
which, by analogy with the hydrodynamics \cite {Batchelor}, can be called the vorticity of the electromagnetic field. Sometimes this term is used to refer to the curl of the Poynting vector of a vortex laser beam \cite {Berry2009}.
The components of the vector $ \bm \Omega $ designate the torque acting on the area of finite size, orthogonal to the corresponding axis. In the experiment with a twisted laser beam \cite {ONeil2002}, the rotation of small particles away from the axis of the laser beam was observed. This is probably a manifestation of the nonzero vorticity of the laser beam in this direction.

Speaking of torque acting on the area, we mean the flux of the normal component of angular momentum through the area.
Actually, the angular momentum absorbed by a real object depends significantly on the optical properties of the material of the object and on diffraction at its edges.

 Until now, we have not made any assumptions about the distance between the charge and the point of observation. Let us now find the asymptotic expressions for the torque of the field at distances $ r, R \gg r '$. We represent the vectors $ \bm E $ and $ \bm H $ as the sum of the `large' and `small' components $ \bm E = \bm E_1 + \bm E_2 $ and $ \bm H = \bm H_1 + \bm H_2 $. We introduce a  vector $ \bm g _ {\perp}$ with components $ \sigma_ {i1} $. The vector $ \bm g _ {\perp} $  is  orthogonal to the vector $ \bm n $ and represents the shear force acting tangentially on the area orthogonal to the vector $ \bm n $. 
 Arguing as in the derivation of Eq. (\ref {Lds1}), we obtain a similar expression for the vector $ \bm g _ {\perp} $  in a spherical coordinate system
 \be\label{s11}
\bm g _ {\perp} =-\frac{1}{4\pi}\left[\bm E_1 (\bm E_2\,\bm  n)+H_1(\bm H_2\bm n)\right],\quad i=\theta,\varphi.
\ee
Bearing in mind that in this approximation $ \bm H_1 = (\bm n \times \bm E_1) $ and taking into account the equality (\ref {pper}), we obtain
 \[
\bm g _ {\perp}=\frac 1c (\bm n\times (\bm P\times \bm n)).
 \]
 One can  see from the last expression that the vector $ \bm g _ {\perp} $ is proportional to the transverse, with respect to $ \bm n $, component of the Poynting vector.
Therefore, the lines shown in Fig. \ref {circ} can be interpreted as the field lines of the transverse component of the Poynting vector. Accordingly, the vorticity of radiation in the far-field approximation is
\be \label {Om1}
\Omega_n = \frac {1} {c} (\bm n \rot \bm P).
\ee
Eqs. (\ref {Lds-n}), (\ref {Lds-n1}) and (\ref {Om1}) show that in the far-field approximation, the angular momentum flux is determined by the transverse component of the Poynting vector, and the torque, acting on the area of finite dimensions is proportional to the curl of the transverse component of the Poynting vector.
 
\section{Rotating magnetic dipole\label{VI}}

\begin{figure}[ht]\center
\includegraphics[width=4cm]{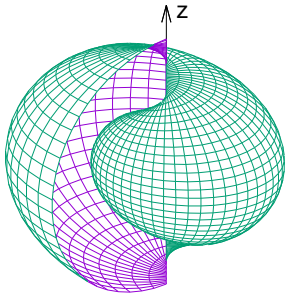}
\caption{The wavefront of the field defined by equations (\ref {muEH}) has typical singularity on the $ z $ axis.}
\label{phase}
\end{figure}

As a simple example, consider the field produced by a rotating magnetic dipole. We chose a stationary field source as an example to avoid significant complication of calculations associated with the dependence of the field on the retarded time. The law of motion of the dipole  vector $ \bm \mu $ in the Cartesian coordinate system ($ x, y, z $) is 
\begin{equation}
\label{mut}
\bm\mu=\mu(\cos\omega t,\sin\omega t,0),
\end{equation}
where $\mu$ and $\omega$ are the modulus and angular velocity of rotation of the dipole. The  electromagnetic field in a spherical coordinate system is as follows
\begin{align}
E_\theta=&-\frac{\mu\omega}{r^2 c}(\cos\tau-\rho\sin\tau),\nn\\
E_\varphi=&-\frac{\mu\omega}{r^2c} \cos \theta (\sin\tau+\rho\cos\tau)\nn,\\
H_r=&\frac{2\mu}{r^3} \sin\theta (\cos\tau-\rho\sin\tau),\label{muEH}\\
H_\theta=&-\frac{\mu}{r^3}\cos\theta (\cos\tau-\rho\sin\tau-\rho^2\cos\tau),\nn\\
H_\varphi=&-\frac{\mu}{r^3} (\sin\tau+\rho\cos\tau-\rho^2\sin\tau),\nn
\end{align}
where
\[
 \rho= \frac{\omega r}{c},\quad \tau=\omega t'-\varphi,\quad t'=t-\frac  rc.
\]
This field obviously has the properties of a vortex field. The wavefront given by the equation $ \tau = $ const is shown in Fig. \ref {phase}.
  
\begin{figure}[ht]\center
\includegraphics[width=4cm]{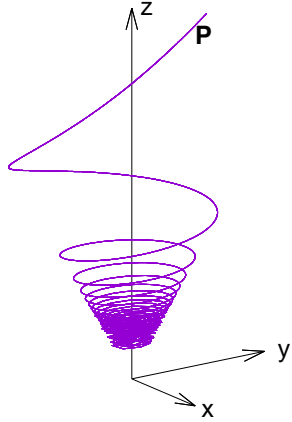}
\caption{Poynting vector force line}
\label{poynt}
\end{figure}

The Poynting vector has the components
\begin{align}\label{Poyt}
P_r=&P_0[\rho ^3  ( \sin ^2\tau +\cos ^2\theta \cos ^2\tau)-\rho ^2  \sin 2\tau \sin ^2\theta +\rho   \cos
 2\tau \sin^2 \theta + \sin \tau \cos\tau\sin ^2\theta],\nn \\
P_\theta =&P_0 \sin \theta\cos \theta  (\rho ^2\sin 2\tau
-2\rho \cos 2\tau -\sin 2\tau ) ,\\
P_\varphi=&2P_0 \sin \theta \left( {\rho ^2\sin ^2\tau -\rho \sin
2\tau +\cos ^2\tau } \right),\nn
\end{align}
where
\[
P_0=\frac{\mu^2\omega}{4\pi r^5}.
\]
After averaging over time we obtain
\be\label{Pav}
\langle\bm P\rangle=\frac{\mu^2\omega^7}{8\pi c^6 \rho^5}\left[\rho^3(1+\cos^2\theta)\bm e_r+2(1+\rho^2)\sin\theta\,\bm e_\varphi\right].
\ee
The last equation shows that the Poynting vector is directed tangentially to the surface of a cone with apex at the origin and an angular opening $ \theta $. Fig. \ref {poynt} shows one of the lines of force of the Poynting vector.
In the near zone, in the region of small $ \rho $, the azimuthal component $ \langle P_ \varphi \rangle $ prevails, and in the far zone prevails the radial component $ \langle P_r \rangle $, which determines the radiation intensity and decreases as $ 1 / r ^ 2 $. The orthogonal to $\bm r$ component, responsible for the angular momentum, is relativelly large in the near field region and falls off as $1/r^3$ in the far field region.

The angular momentum flux is determined by the off-diagonal elements of the stress tensor
\begin{align}
\sigma_{12}=&\frac{\mu^2}{4\pi r^6}\sin 2\theta \left(\cos ^2\tau-\rho \sin 2\tau -\rho^2 \cos 2\tau +\frac 12 \rho^3 \sin 2\tau\right),\nn\\
\sigma_{13}=&\frac{\mu^2}{4\pi r^6}\sin \theta(\sin 2\tau+2\rho\cos2\tau-2\rho^2\sin 2\tau+2\rho^3 \sin^2\tau ) ,\\
\sigma_{23}=&-\frac{\mu^2}{4\pi r^6} \cos \theta \left( \frac 12\sin 2\tau +\rho \cos 2\tau -\rho^2\sin 2\tau\right),\nn
\end{align}
Substituting this into the Eq. (\ref {Lds-sp}) and averaging over time, we obtain the angular momentum flux in the radial direction at any distance from the dipole
\be\label{Ldip2}
\left\langle\frac{\d\bm L}{\d t\d s}\right\rangle=\frac{\mu^2\sin\theta}{4\pi r^5}\left(\bm e_\varphi\cos\theta-  \bm e_\theta\rho^3\right).
\ee
At larger distances, only the $\theta$-component of flux remains in  radiation
\be\label{Ldip1}
\left\langle\frac{\d\bm L}{\d t\d \Omega}\right\rangle=-\frac{\mu^2\omega^3}{4\pi c^3}\sin\theta\,\bm{e}_\theta.
\ee
Note, that the angular momentum flux has its maximum in direction $ \theta = \pi / 2 $ and is  zero on the axis of rotation. Eq. ({\ref{Ldip1}) can be obtained by use of  Eq. (\ref {Lds-n1}) if the Poynting vector (\ref {Pav}) is averaged over time and the term with the highest power of $ r $ remains. However, at arbitrary distances Eq. (\ref {Lds-n1}) is not correct.

The integral of the expression (\ref {Ldip1}) over the solid angle gives the well-known formula for the rate of loss of the angular momentum of the dipole due to radiation reaction
\[
\left\langle\frac{\d\bm L}{\d t}\right\rangle=\frac{2\mu^2\omega^3}{3 c^3}\bm{\hat z}.
\]

In order to find the vorticity of the electromagnetic field, we calculate the radial component of the vector $ \bm \Omega $. According to Eq. (\ref {omn}), we have
\be\label{r-4}
\Omega_n=\frac{1}{r\sin\theta}\left(\frac{\partial (\sigma_{31}\sin\theta)}{\partial \theta}-\frac{\partial \sigma_{21}}{\partial \varphi}\right)=\frac{\mu^2\omega^3}{2\pi c^3 r^4} \cos \theta.
\ee
This value, in contrast to the angular momentum flux, is maximum in the direction of the $ z $ axis, that is, in the direction of the vortex axis (see Fig. \Ref {phase}). In the equatorial plane $ xy $, the vorticity is zero. This property resembles the polarization of radiation. It can be seen from Eqs. (\ref {muEH}) that in the  $ z $ axis direction  the field is circularly polarized, and in the equatorial plane it is linearly polarized, since
\[
\frac{E_\varphi^2}{\cos^2\theta}+E_\theta^2=\text{const}.
\]

The sign of $ \Omega_n $ changes when passing through the equatorial plane - the sign of vorticity coincides with the sign of the projection of the angular velocity of rotation of the dipole onto the direction of radiation. We also note that $ \Omega_n $ decreases with distance as $ 1 / r ^ 4 $ and this dependence is the same at both large and small distances from the dipole.

Finally, we calculate the radial component of the  curl of the  Poynting vector. Using Eqs. (\ref{Poyt}) and averaging over time we obtain
\[
\langle(\bm n\rot\bm P)\rangle=\frac{\mu^2\omega^3}{2\pi c^2 r^4} \cos \theta (1+\rho^{-2}).
\]
At large distance this coincides with  the vorticity of radiation $\Omega_n$ according to Eq.  (\ref {Om1}).

\section{Discussion\label{VII}}
We investigated the angular momentum of an electromagnetic field of an arbitrarily moving point charge. In the general case, the  angular momentum flux is determined by the stress tensor of the electromagnetic field (Eq. (\ref{Lds-sp})). From a practical point of view, the angular momentum flux at large distances from the charge is of interest. Namely, at distances much greater than the characteristic  wavelength of radiation. This area is usually referred to as a wave zone.
We have shown that the flux of angular momentum in the wave zone is proportional to the vector product of the radius vector of the observation point by the Poynting vector (Eqs (\ref {Lds-n}) and (\ref {Lds-n1}). Since the angular momentum depends significantly on the choice of coordinate system, two limiting cases can be distinguished:

i) the distance between the charge and the origin is comparable to the distance between the charge and the point of observation, and

ii) the distance between the charge and the origin is much less than the distance between the charge and the point of observation.

In the first case, the flux of the angular momentum of  radiation can be interpreted as the pressure of radiation
 on a target that produces a torque around the origin of the coordinate system. In this case,  one can take into account only the main part of electromagnetic field  decreasing with distance as $1 / r$ and assume that the vectors of the electric field, magnetic field, and the Poynting vector are mutually orthogonal.

In the second case, it is necessary to take into account the components of the electromagnetic field that decrease with distance as $ 1 / r ^ 2 $. Thus, we take into account that the direction of the Poynting vector does not coincide with the direction of radiation, more precisely, with the direction of the radius vector of the observation point. In this case, the flux of the angular momentum of the radiation is determined by the component of the Poynting vector that is transverse to the radius vector.
In this last case, we obtained an explicit expression for the flux of angular momentum of radiation as a function of coordinates, velocity and acceleration of the charge (formula (\ref {vec})).

For practical applications, an important property of radiation is the torque acting on the object due to electromagnetic field. For example, to use radiation as optical tweeze rs and optical traps \cite {He1995, Simpson1997, Zhao:09, Bruce2021}. From this point of view, it is of interest to find the flux of the radial component of the angular momentum through an area of finite  dimensions located perpendicular to the radius vector.
We have shown that this quantity, denoted as the vorticity of the radiation, is proportional to the curl of the stress tensor (Eq. (\ref {bm-Om})), and in the wave zone it is proportional to the curl of the Poynting vector (Eq. (\ref {Om1})).

\begin{acknowledgments}
The research was supported  by Ministry of Science and Higher Education of Russian Federation, project FEWF-2020-0003; and by RFBR project No. 19-42-700011. 
\end{acknowledgments}


%

\end{document}